# “双子”，中子与中子星

## ——漫谈朗道与中子星

徐仁新

北京大学物理学院，北京 100871

**摘要:** 中子与中子星这一话题在近代物理或原子物理中多有涉及,但往往存有误解。本文回顾并评述了中子的发现及朗道提出中子星概念的过程,试图将这一历史准确全面地反映出来。

中子星因其极端物理环境而深受物理学和天文学领域学者们的青睐,其研究在这两个领域均不可或缺。天文学家观测发现的若干类高能天体现象与中子星的存在紧密相关，而物理学家则将中子星当作重要的天体实验室来认识自然基本规律。值得一提的是，中子星概念最早由朗道提出，历史上可以追溯到 1930 年代初中子的发现。对这段脍炙人口历史的回顾，不仅使人因身临其境地体会“微观”与“宇观”的学科交融而受益，而且有助于培养科学创新精神。这一题材在基础物理教育中显然具有现实意义。

前苏联理论学家朗道（L. Landau）是公认的物理全才和奇才。他 1908 年出生，1962 年因严重车祸而丧失研究能力并于同年“因他对于凝聚态物质（特别是液氦）的先驱性理论”而荣获诺贝尔物理学奖，1968 年去世。在教育方面，朗道与栗弗席兹等筹划和编写的十卷本巨著《理论物理教程》对国际理论物理的教育影响深远，至今仍被奉为“经典”。

关于朗道生平、学术传记式的介绍很多（如[1]，近期如文献[2]及其中引文），且其传奇式的“八卦”故事也遍及网络（读者不妨通过“Google”或“百度”搜索）。然而这些作者们大多侧重于朗道作为一个物理学家，特别是凝聚态物理学家的一面，而鲜少涉及他对天体物理，特别是“中子星”概念的提出的另一面。比如，人们通常认为朗道最重要、最著名的贡献是概括在所谓的“朗道十诫[2]”中，而那里就未包含“中子星”概念的提出。本文拟弥补这一缺憾。

在 1937 年苏联政治气氛紧张期间，朗道为了将自己塑造成包括西方在内的全世界认可的学术权威以抵消自己的社会压力，他向当时著名杂志 *Nature* 投稿一篇题为“恒星能量的起源”的文章（其译文见本文附录）。尽管此文 1938 年 2 月发表于 *Nature*[3]，但终究未能改变朗道当年 4 月 28 日被捕入狱之噩运（在 Kapitsa 等人一再请求和担保下，才于次年 4 月 29 日出狱）。据文献[4]（该文题为“Complete list of L D Landau’s works”），朗道一生共发表了 Nature 杂志论文六篇，其中独立署名的三篇 Nature 文章就包括这篇关于恒星能源的短文（另外两篇独立署名论文是分别总结他之前关于相变和超流理论的）。读者或许由此评估那篇 Nature 短文在朗道心目中的份量；但值得注意的是，该短文中提及的关于中子星和恒星能源的概念其实早已出现于他 1932 年写的一篇较详细的论文[5]（译文见文献[6]的“附录 1”）。由此可见，不应轻易忽略朗道作为一名天体物理学者的贡献。

总之，应该说朗道是比较在意自己关于恒星能源和中子物质的想法的。本文试图介绍并讨论朗道的天体物理这一面，以对众多“朗道读物”作一补充。为此，让我们从十九世纪末、二十世纪初那个激动人心的时代开始。

**从一个古老的概念 ——“双子”谈起** 相对于汤姆孙1897年发现电子而言，中子的发现要晚得多。中子概念的提出是与人类探索原子核（英国实验物理学家Rutherford根据α散射实验提出原子的“有核模型”）组成的过程分不开的。1920年Rutherford在题为“原子核组成”的讲座中强调到了实验上发现一些原子核的电荷数约只有其质量数的一半，并指出[7]：“…

电子可能与氢核很紧密地结合而形成一类新的中性双子（doublet）”。他还推测双子会在物质中自由运动，难被探测，或许不能被限制于密封容器中。人们一般认为，1921年Harkins在讨论同位素分类时明确地用“中子”（neutron）这个词代替Rutherford的“双子”。

中子的发现也曾经历一番周折。1930 年 Bothe 和 Becker[8]报道用 Po 的α射线轰击 Be 时产生的一种中性“γ射线”，它比一般γ射线的穿透能力强很多。1932 年 Joliot-Curie 夫妇[9]用这种“γ射线”轰击石蜡中的质子，并且根据出射质子的能量推算“γ射线”的能量（约为50MeV）。为何 Be 原子核被α射线轰击后能够产生如此高能的γ射线？当年，深受 Rutherford “双子”概念影响的 Chadwick 注意到 Joliot-Curie 夫妇的实验，并怀疑这种射线不是γ射线而是他一直实验寻找的中子（“双子”）。为了搞清楚 Bothe 等发现的射线到底是γ射线还是中子，Chadwick 将这种射线轰击 He、Li、N 等其他原子核，也测得了这些核不同的反冲动能。如果认为那种射线是零静止质量的γ射线，则根据反冲动能推算的γ射线能量差别很大；但若认为那种射线就是质量跟质子相近的中子，就没有这些矛盾了。因此 Chadwick 认为“中子可能存在”[10]，并因中子发现很快荣获 1935 年度 Nobel 物理学奖。

**“密度跟原子核相当的物质”概念的提出** 英国核物理学家 Chadwick 发现中子后不久，Iwanenko[11]等人相继提出原子核由质子和中子组成。Iwanenko（1904～1994）是朗道在列宁格勒大学本科时期的同学。在本科阶段，他俩与另一位同学 Gamow（1904～1968）曾经一起学习和评论相对论和量子论，受人瞩目，时称少年“三剑客”；也曾一起合作发表过研究论文。Gamow 是热大爆炸宇宙学、α衰变的势垒隧穿机制以及氨基酸三联码概念的提出者。可以想象，中子的发现或许也曾经是这三人热闹讨论的话题。Iwanenko 的这篇文章很短，全文复制如下*。

**The Neutron Hypothesis**

Dr. J. Chadwick's explanation [1] of the mysterious beryllium radiation is very attractive to theoretical physicists. Is it not possible to admit that neutrons play also an important rôle in the building of nuclei, the nuclei electrons being *all* packed in α-particles or neutrons? The lack of a theory of nuclei makes, of course, this assumption rather uncertain, but perhaps it sounds not so improbable if we remember that the nuclei electrons profoundly change their properties when entering into the nuclei, and lose, so to say, their individuality, for example, their spin and magnetic moment.

The chief point of interest is how far the neutrons can be considered as elementary particles (something like protons or electrons). It is easy to calculate the number of α-particles, protons, and neutrons for a given nucleus, and form in this way an idea about the momentum of nucleus (assuming for the neutron a moment $\frac{1}{2}$). It is curious that beryllium nuclei do not possess free protons but only α-particles and neutrons.

D. Iwanenko.

Physico-Technical Institute,
Leningrad, April 21.

Nature, **129**, 312, Feb. 27, 1932.

---

* 有证据显示“三剑客”对于原子核组成都有兴趣。在发现中子之前的 1928 年，Gamow（*Z. Physik*, 1928, **51**, 204）就提出α衰变的α粒子量子遂穿机制；而在发现中子之后不久，Landau 就探讨了极端相对论电子与质子作用形成致密“中子物质”的可能性（例如，见本文附录）。除 Iwanenko 此文，Heisenberg（*Zeits. Fur Phys.*, 1932, 77, 1）也独立地提出原子核由质子、中子组成的概念。不过，Sergey Bastrukov 在北大访问期间告诉我：坊间盛传是 E. Gapon 告诉 Iwanenko “中子”之后才成此文的。Gapon 和 Ivanenko 曾于 1932 年提出第一个原子核壳层模型（http://www.g-sardanashvily.ru/IvanenkoGapon.pdf）。目前这“三剑客”均已作古，但“盖棺定论”式的评价不一。Iwanenko 被传是将 Landau 送进监狱的重要推手；Gamow 三大贡献中任意一个都是 Nobel 奖量级的，他被做出重要成就但未获诺奖者当作典范而以求安慰。有趣的是 Iwanenko 等 1965 年提出中子星内部可能出现夸克物质，后来学界在此看法基础上还发展出“夸克星”的概念；而观测发现的脉冲星到底是中子星还是夸克星，至今乃在争论之中（参见：徐仁新和岳友岭，2006，科技导报，24 卷 5 页）。“三剑客”故事充分应验了古语：“古今多少事，都付笑谈中”。

该文先肯定了 Chadwick 对 Be 射线的解释，再提出所有“核电子”（nuclei electrons）挤进原子α粒子或中子并丧失电子原有属性（如自旋和磁矩等）的可能性，更指出中子多大程度上能像质子和电子那样当作基本粒子。这些观点是难能可贵的；要知道，第一个弱相互作用的定量理论是 Fermi 在 1934 年才完成的[12]。

朗道推测存在核密度物质甚至早于 Chadwick 发现中子。1932 年 1 月 7 日“Physikalische Zeitschrift der Sowjetunion”杂志社收到朗道的一篇稿件*，后被接受发表于该刊[6]。目前一般认为，有关“中子星”概念的原型就是在这篇论文中首次提出的；1938 年的论文[3]是该文的总结和深化。朗道提出“中子星”概念的逻辑思路很值得科学史学者参考；概括于如下两点。

A，首次提出可能存在核心由“密度跟原子核相当的物质”组成的星体（后称“中子星”）。这一点具有前瞻性，而中子星逐渐成为天文学和物理学领域的研究热点。类似于 Iwanenko 对于原子核结构的思索，朗道的观点是，星核足够高的密度不仅可以因引力能的下降导致整个星体稳定，而且一旦质子和电子“紧密结合”，将使得电子的动能下降。关于恒星的整体结构，朗道认为恒星的“中子核心”与周围物质“这两个相的边界条件是由通常的化学势平衡确定”；而任一现代的中子星结构正是沿此观点建立的。

B，认为利用这种星体结构可以自然地解决当时流行的两个挑战性问题：恒星塌缩和能源机制。如果只有以上那一点“A”的话，朗道中子星方面的探索或许仅为又一“屠龙绝技”，因为当时的天文观测方面不存在这样一种恒星结构的必要。所以，为了将他的中子星概念在学术界“卖”出去，朗道声称他的星体结构可同时解决“恒星塌缩”和“能源机制”疑难**。为此，朗道不惜提出了现在看来是错误的两个概念：一、所有恒星都具有一个“中子核心”以便提供发光能源；二、恒星内部存在“违背量子力学定律的病态区域”以便使质子和电子“紧密结合”得像基本粒子那样（如果量子力学定律成立的话，质子和电子只能结合成过大的“氢原子”）。

应该说，以上错误的“B”对于合理的“A”在学术界的流行是有帮助的。我们今天很难考证当年朗道是先想到了“A”还是“B”，但有一点是清楚的：朗道对当时“恒星塌缩”和“能源机制”两个疑难问题了解得全面、深入。可见，科学的进步终由问题所驱动。

胡适曾言“大胆假设、小心引证”以教诲人们如何拿捏“创新”和“严谨”，但这一哲学在实际研究工作中却很难精确地指导。尽管朗道的这两篇论文[3,6]中充斥着不少经不起时代推敲的错误概念，但七十余年后的今天我们还不得不提及它们；因为“密度跟原子核相当的物质”概念是在那里首先提出的，并且为了理解当今探测到的大量天文观测现象，我们离不开这种物质。

**致谢**：承蒙陆埮先生阅读初稿并一起讨论、回顾朗道与中子星这段历史。作者受益匪浅，谨此感谢。

* 朗道于 1931 年 2 月访问苏黎士时撰写此文。Chadwick 于 1932 年 2 月 24 日写了一封信给 Bohr，讨论中子的发现。

** 事实上，在中子发现之前，不少著名物理学者就已经非常关心恒星引力平衡和塌缩的问题了。这是当时的一个重要研究背景。另外一个背景是当时不清楚恒星的产能机制。

## 附录："恒星能量的起源"

（译自：L. Landau, 1938, Nature, 141, 333）

众所周知，物质由原子核和电子组成。然而我们能够看到，在非常重物体的内部，物质的这一常规"电性"状态是不稳定的。个中的原因源于如下事实：物质的"电性"状态不能存在于密度极端高的情形，因为在如此高密度下电子形成费米气、具有极高的压强。另一方面，很容易看出物质可能进入另外一个更致密的状态——那里所有的原子核跟电子结合在一起形成中子[1]。即使假设中子间互相排斥，这一斥力也只有在密度达到核密度量级（~$10^{14}$ g/cm$^3$）时才显著；因中子质量较高，中子费米气的压强远低于同样密度下的电子费米气压强。

尽管由于形成中子的反应是吸热的，通常条件下物质的"中子"状态并非满足能量极低，但这一状态在物体的质量足够大时却还是稳定的。在后一中情况下，支撑高密度的中子状态而获得的引力能弥补了内能的不利。

很容易计算当"中子"状态开始比"电性"状态更稳定时物体的临界质量。首先我们要计算形成一个中子所必须的能量。例如，考虑反应 ${}^{16}_{8}\mathrm{O} + 8e^- = 16\,{}^{1}_{0}n$，从质量亏损我们发现形成一个中子需要 0.008 倍质量单位（或 $1.2\times10^{-5}$ erg = 7.5 MeV）。将 1g 物质转变成中子我们因而需要消耗 $7\times10^{18}$ erg 的能量。

现在我们要计算所获得的引力能。不太致密的"电性"状态的引力能当然是可以忽略的。让我们先假设中子状态具有常密度 $10^{14}$ g/cm$^3$。质量为 $M$ 的均匀球体的引力能是 $3\times10^{-3}\,M^{5/3}$ erg。为了得到稳定的中子相，我们必须有：$3\times10^{-3}\,M^{5/3} > 1.2\times10^{-5}\,M$，或 $M > 10^{32}$ g = 0.05 $M_\odot$；这里 $M_\odot$为太阳质量。另一方面，如果我们假设中子的行为类似费米气体，具有能量 $7\times10^{-22}\,M^{7/3}$ erg，因此有

$$M > 1.5\times10^{30}\ \mathrm{g} = 10^{-3}\,M_\odot。$$

这一临界值甚至低于依赖前一假设所得到的值。

当物体的质量大于那个临界质量时，则在形成"中子"相过程中将释放巨大的能量，并且我们看到：物质"中子"状态的概念为恒星能源问题给出了一个直接的答案。太阳在其可能的辐射阶段间（根据广义相对论，持续大约 $2\times10^9$ 年）应该发射了量级 $3\times10^{50}$ erg 的能量。释放这些能量只需要太阳约 2%（在常密度假设前提下）或甚至仅 $3\times10^{-3}\,M_\odot$（在费米气模型中）的物质转化为"中子"相。即便如猎户座β那样的亮星，我们发现其中子核心的质量也只有 0.1 $M_\odot$（在费米气模型中）。

我们因而认为恒星应该具有一个中子核心，其稳恒的增长释放着维持恒星高温的能量；这两个相的边界条件还是由通常的化学势平衡确定。对这一模型的详细研究，将可能建立起一个自洽的恒星理论。

至于初始核心是如何形成的问题，我已经阐述[2]：当然必须在质量超过 1.5 $M_\odot$的星体内

部形成这样的一个核心。在质量比较小的恒星内部，形成可能的初始核心的条件尚不清楚。

[1] Hund, F., *Erg. d. exakten Natwis.* 15 189 (1936).
[2] Landau, L., *Sov. Phys.*, 1, 285 (1932).

# "Doublet", Neutron and Neutron stars
## --- An essay on Landau and Neutron stars

**ABSTRACT** The concept of extremely dense matter at supra-nuclear density was first speculated by L. Landau in the beginning of 1930s when neutron was just discovered. A historical review on these issues not only explains the interaction between micro and cosmic physics, but also has profound implications for scientific innovation. It is surely meaningful in realistic physics education to look back to this history. (The review was published in Chinese.)